\documentstyle[aps,prd]{revtex}
\tighten
\renewcommand{\epsilon}{\varepsilon}
\begin{document}

\title{Reacting fluids in the expanding Universe:
A new mechanism for entropy production}

\author{Winfried Zimdahl\footnote{Electronic address:  
winfried.zimdahl@uni-konstanz.de}}

\address{Fakult\"at f\"ur Physik, Universit\"at Konstanz, PF 5560 M678
D-78434 Konstanz, Germany}

\date{February 7, 1997}

\maketitle

\begin{abstract}
It is shown that two reacting cosmological fluids,
each of them perfect on its own, which exchange
energy and momentum without preserving particle
numbers, give rise to an entropy producing `reactive' bulk
stress of the system as a whole, as soon as the
detailed balance between decay and inverse decay
processes is perturbed.
This demonstrates explicitly that particle generation
is dynamically equivalent to an effective bulk pressure.
We derive a semiquantitative formula for the corresponding
new kinetic coefficient and evaluate the latter for the
out-of-equilibrium decay of heavy, nonrelativistic particles into  
radiation.
It turns out that the associated reactive bulk viscosity may be  
more than
one order of magnitude larger than the conventional bulk viscosity, 
calculated, e.g., in radiative hydrodynamics.
\end{abstract}


\section{Introduction}
Understanding the origin of the entropy content of the Universe is one 
of the major problems in cosmology.
Entropy production requires deviations from thermodynamical equilibrium. 
Such deviations  typically occur if microscopic interaction time scales 
become comparable with cosmologically relevant time scales.
Phenomenologically, entropy producing processes may be described with 
the help of dissipative fluid models.
From the standard dissipative fluid effects, heat conductivity, shear 
viscosity and bulk viscosity, the latter is the most favoured one  
in a cosmological context since it is the only effect which is  
compatible
with the symmetry requirements of the homogeneous and isotropic  
Friedmann-Lema\^{\i}tre-Robertson-Walker (FLRW) universes.
The existence of a nonvanishing bulk viscosity for a simple relativistic 
gas has been known since the work of Israel \cite{I}.
A radiative bulk viscosity coefficient was derived by Weinberg  
\cite{Wein}
who also gave an analysis on its role in cosmology, followed by  
subsequent investigations by Straumann \cite{Strau}, Schweizer  
\cite{Schw} and Udey
and Israel \cite{UI}.
Furthermore, bulk pressures were shown to play a role in  
quark-hadron transitions \cite{DaGyu}.
It was explicitly demonstrated recently \cite{ZMN} that the  
occurence of a cosmological bulk viscosity is a general feature of  
any system of
interacting fluids with different cooling rates in the expanding
universe.
The considerations of all the mentioned papers were based on the
assumption that the particle numbers of each of the components are
separately conserved during the interaction process.
But particle number conservation is only a very special case, especially 
at high energies.
In the present paper we consider a two-fluid universe with mutually 
interacting
and reacting components
that are allowed to exchange energy and momentum and to
convert into each other.
The corresponding change in the number of fluid particles
is modelled by simple rate equations.
In general, these reactions do not preserve the total
number of fluid particles.
Each of the interacting and reacting fluids is
assumed to be perfect on its own.
As long as the components are in thermal equilibrium and
the conversion rates obey detailed balance relations, the
entropy production of the system as a whole vanishes.
Equilibrium situations like these are assumed to be realized
at very high temperatures when the equations of state of
both of the components and of the system as a whole are
those for relativistic matter.
As the Universe cools down, one of the subsystems, say
fluid $1$, becomes nonrelativistic while the other fluid,
say fluid $2$, continues
to behave like radiation.
The equations of state of the fluids are now different.
Moreover, the detailed balance relations are no longer
satisfied in general.
The reactions may proceed preferentially in one direction,
the inverse processes being suppressed.
Typical cases for this kind of situation   are the
out-of-equilibrium decay of heavy bosons into  quarks and
leptons during the process of baryogenesis, or the scalar
field decay into relativistic particles
 in the reheating phase of
`standard' inflationary scenarios \cite{KoTu}. Also, the annihilation 
of electrons and positrons into photons at the nucleosynthesis
energy scale belongs to this category.

The purpose of this paper is to demonstrate that
{\it the difference in the equations of state of the fluid
components and the simultaneous deviation from the detailed
balance in the rate equations
for the interfluid reactions  give rise to a new entropy
producing cosmological effect that manifests itself as a
`reactive' bulk pressure}.
We shall derive here a semiquantitative formula for this
effect for
small deviations from equilibrium and apply it
to a typical out-of-equilibrium situation in the early universe.

To point out the difference to conventional bulk pressures,
calculated, e.g., in radiative hydrodynamics, we
emphasize that the physical reason for the effect under
consideration here is twofold.
The first reason is that because of their different
equations of state  the fluid subsystems are subject
to different cooling rates in the expanding Universe.
The second reason are the deviations from the equilibrium
balances
between the decay and inverse decay reactions of the fluids.
It is the combination of both these causes that produces
the effect to be discussed in the present paper.
For conventional bulk viscosities, differences in the
cooling rates alone are sufficient \cite{UI,ZMN}.

The described effect implies that in a reacting mixture
the generation of particles is dynamically equivalent
to an effective viscous pressure.
For a decay rate comparable with the expansion rate we
show this effect to be more than one order of magnitude
larger than the corresponding effect due to a
conventional bulk viscosity.

The structure of the paper is as follows:
Section 2 establishes the basic theory of two interacting
and reacting fluids with generally different temperatures.
Assuming that the interactions establish an approximate
equilibrium between the fluids,
we investigate the same system within an effective
one-temperature model in section 3.
Comparing both descriptions, specifically the respective
expressions for the entropy production, we derive a general,
semiquantitative formula for the reactive bulk
pressure of the system as a whole in section 4.
Section 5 presents an application of this formula
to a  mixture of a
nonrelativistic and a relativistic fluid and compares
the reactive bulk viscosity coefficient with the conventional one.
A brief summary of the paper and a few comments on
cosmological applications are given in section 6. \\
Units have been chosen so that $c = k_{B} = \hbar = 1$.
\section{The two-fluid model}
The energy-momentum tensor $T ^{ik}$ of the cosmic
substratum is assumed to split into  two perfect fluid parts,
\begin{equation}
T ^{ik} = T ^{ik}_{_{1 }}
+ T ^{ik}_{_{2 }} {\mbox{ , }}
\label{1}
\end{equation}
with ($A = 1, 2$)
\begin{equation}
T^{ik}_{_{A }} = \rho_{_{A }} u^{i}u^{k} 
+ p_{_{A }} h^{ik}{\mbox{ .}} \label{2}
\end{equation}
$\rho_{_{A }}$ is the energy density and 
$p_{_{A }}$ is the equilibrium pressure of species $A$.  
For simplicity we assume that both components share the
same $4$-velocity $u^{i}$. The quantity
$h^{ik}$ is the
projection tensor 
$h^{ik} =g^{ik} + u^{i}u^{k}$. 
The particle flow vector $N_{_{A }}^{i}$ of species $A$ is defined as 
\begin{equation}
N_{_{A }}^{i} = n_{_{A }}u^{i}{\mbox{ , }}
\label{3}
\end{equation}
where $n _{_{A }}$ is the particle number density.
We are interested in situations where neither the particle
numbers nor the energy-momenta of the components are
separately conserved, i.e.,
conversion of particles and exchange of energy and momentum
between the components are  admitted.
The balance laws for the particle numbers are
\begin{equation}
N _{_{A } ;i}^{i} = \dot{n}_{_{A }}
+ \Theta n_{_{A }} = n _{_{A }} \Gamma _{_{A}}
{\mbox{ , }}
\label{4}
\end{equation}
where $\Theta \equiv u^{i}_{;i}$ is the fluid expansion
and $\Gamma _{_{A}}$ is the rate of change of the number of
particles of species $A$.
There is particle production for $\Gamma _{_{A}} > 0$ and
particle decay for
$\Gamma _{_{A}} < 0$, respectively. For $\Gamma _{_{A}} = 0$,
we have separate particle number conservation.

Interactions between the fluid components amount to
the mutual exchange of energy and momentum.
Consequently, there will be no local energy-momentum
conservation for the subsystems separately.
Only the energy-momentum tensor of the system as a
whole is conserved.

Denoting the loss- and source-terms in the separate
balances by $t ^{i}_{_{A }}$,
we write
\begin{equation}
T ^{ik}_{_{A } ;k} = - t _{_{A }}^{i} {\mbox{ , }}
\label{9}
\end{equation}
implying
\begin{equation}
\dot{\rho}_{_{A }}
+ \Theta\left(\rho_{_{A }}
+ p_{_{A }}\right)
= u _{a} t _{_{A }}^{a}
{\mbox{ , }}
\label{10}
\end{equation}
and
\begin{equation}
\left(\rho _{_{A }}
+ p _{_{A}}\right) \dot{u}^{a}
+ p _{_{A } ,k}h ^{ak}
= - h ^{a}_{i}t ^{i}_{_{A }} {\mbox{ .}}
\label{11}
\end{equation}
All the considerations to follow will be independent
of the specific structure of the $t _{_{A }}^{i}$.
In other words, there are no limitations on the strength
or the structure of the interaction.

Each component is governed by a separate Gibbs equation
\begin{equation}
T _{_{A }} \mbox{d} s _{_{A }}
= \mbox{d} \frac{\rho _{_{A }}}{n _{_{A }}}
+ p _{_{A }}
\mbox{d} \frac{1}{n _{_{A }}} {\mbox{ , }}
\label{12}
\end{equation}
where $s _{_{A }}$ is the entropy per particle of species $A$.
Using eqs.(\ref{4}) and (\ref{10}) one finds for the time
behaviour of the entropy per particle
\begin{equation}
n _{_{A }} T _{_{A }}
\dot{s}_{_{A }} = u _{a} t _{_{A }}^{a}
- \left(\rho _{_{A }}
+ p _{_{A }}\right) \Gamma _{_{A}} {\mbox{ .}}
\label{13}
\end{equation}
With nonvanishing source terms in the balances for
$n _{_{A }}$ and $\rho _{_{A }}$, the change in the
entropy per particle is different from zero in general.
Below we shall deal with the special case that the terms
on the right-hand side of eq.(\ref{13}) cancel.

The equations of state of the fluid components
are assumed to have the general form
\begin{equation}
p_{_{A }} = p_{_{A }}
\left(n_{_{A }}, T_{_{A }}\right) \label{14}
\end{equation}
and
\begin{equation}
\rho_{_{A }} =
\rho_{_{A }} \left(n_{_{A }},
T_{_{A }}\right){\mbox{ , }}
\label{15}
\end{equation}
i.e.,  
particle number densities $n_{_{A }}$ and temperatures
$T_{_{A }}$
are regarded as the  basic
thermodynamical variables.
The temperatures of the fluids are different in general.

Differentiating  relation (\ref{15}), using the balances 
(\ref{4}) and (\ref{10}) as well as 
the general relation
\begin{equation}
\frac{\partial \rho_{_{A }}}{\partial n_{_{A }}} = 
\frac{\rho_{_{A }} + p_{_{A }}}
{n_{_{A }}} 
- \frac{T_{_{A }}}{n_{_{A }}}
\frac{\partial p_{_{A }}}
{\partial T_{_{A }}} {\mbox{ , }}
\label{16}
\end{equation}
that follows from the requirement that the entropy is a state
function, 
we find the following expression for the temperature
behaviour \cite{Calv,LiGer,ZPRD}:
\begin{equation}
\dot{T}_{_{A }}  = - T_{_{A}} \left(\Theta -
\Gamma _{_{A}} \right)
\frac{\partial p_{_{A}}/\partial T_{_{A}}}{\partial \rho_{_{A}}/
\partial T_{_{A}}}
+ \frac{u _{a} t _{_{A}}^{a} - \Gamma _{_{A}} \left(\rho _{_{A}}
+ p _{_{A}}\right)}
{\partial \rho _{_{A}}/ \partial T _{_{A}}}
{\mbox{ .}}
\label{17}
\end{equation}
Both the source terms $\Gamma _{_{A}}$ and $u _{a}t _{_{A}}^{a}$
in the balances (\ref{4}) and (\ref{10}) backreact on the temperature. 
The numerator of the second term on the right-hand side of eq.(\ref{17}) 
coincides with the right-hand side of eq.(\ref{13}), i.e., the
corresponding
terms disappear in the special case $\dot{s}_{_{A}} = 0$.

For $\Gamma _{_{A}} = u _{a}t _{_{A}}^{a} = 0$ and
with $\Theta = 3\dot{R}/R$, where $R$ is the scale factor of the
Robertson-Walker metric, the equations of state 
$p_{r} = n_{r}kT_{r}$, $\rho_{r} = 3n_{r}kT_{r}$ for radiation
(subscript `$r$')
reproduce the well
known $T_{r} \sim R^{-1}$ behaviour. 
With  $p_{m} = n_{m}kT_{m}$, $\rho_{m} = n_{m}m +
\frac{3}{2}n_{m}kT_{m}$ one recovers $T_{m} \sim R^{-2}$ 
for matter (subscript `$m$'). 

The entropy flow vector $S _{_{A}}^{a}$ is defined by
\begin{equation}
S _{_{A}}^{a} = n _{_{A}} s _{_{A}} u ^{a} {\mbox{ , }}
\label{18}
\end{equation}
and the contribution of component $A$ to the entropy
production density becomes
\begin{eqnarray}
S _{_{A} ;a}^{a} &=& n _{_{A}} s _{_{A}} \Gamma _{_{A}} + n _{_{A}} 
\dot{s}_{_{A}}\nonumber\\
&=& \left(s _{_{A}} - \frac{\rho _{_{A}}
+ p _{_{A}}}{n _{_{A}}T _{_{A}}}
\right)
n _{_{A}}\Gamma _{_{A}} + \frac{u _{a} t _{_{A}}^{a}}{T _{_{A}}}
{\mbox{ , }}
\label{19}
\end{eqnarray}
where relation (\ref{13}) has been used.

According to eq.(\ref{9}) the condition of energy-momentum
conservation for the system as a whole,
\begin{equation}
\left(T _{_{1}}^{ik} + T _{_{2}}^{ik}\right)_{;k} = 0 {\mbox{ , }}
\label{20}
\end{equation}
implies
\begin{equation}
t _{_{1}}^{a} = - t _{_{2}}^{a} {\mbox{ .}}
\label{21}
\end{equation}
There is no corresponding condition, however, for the
particle number balance as a whole.
Defining the integral particle number density $n$ as
\begin{equation}
n = n _{_{1}} + n _{_{2}} {\mbox{ , }}
\label{22}
\end{equation}
we have
\begin{equation}
\dot{n} + \Theta n = n \Gamma {\mbox{ , }}
\label{23}
\end{equation}
where
\begin{equation}
n \Gamma = n _{_{1}}\Gamma _{_{1}} + n _{_{2}}\Gamma _{_{2}}
{\mbox{ .}}
\label{24}
\end{equation}
$\Gamma$ is the rate by which the total particle number
$n$ changes.
We do {\it not} require $\Gamma$ to be zero since total
particle number conservation is only a very special case,
especially at high energies.

The entropy per particle is \cite{Groot}
\begin{equation}
s _{_{A}} = \frac{\rho _{_{A}} + p _{_{A}}}{n _{_{A}}
T _{_{A}}} - \frac{\mu _{_{A}}}
{T _{_{A}}}
{\mbox{ , }}
\label{25}
\end{equation}
where $\mu _{_{A}}$ is the chemical potential of species $A$.
Introducing the last expression into eq.(\ref{19}) yields
\begin{equation}
S ^{a}_{_{A} ;a} = - \frac{\mu _{_{A}}}{T _{_{A}}}
n _{_{A}}\Gamma _{_{A}}
+ \frac{u _{a}t ^{a}_{_{A}}}{T _{_{A}}} {\mbox{ .}}
\label{26}
\end{equation}
For the total entropy production density
\begin{equation}
S ^{a}_{;a} = S ^{a}_{_{1} ;a} + S ^{a}_{_{2} ;a}
\label{27}
\end{equation}
we obtain
\begin{equation}
S ^{a}_{;a} = - \frac{\mu _{_{2}}}{T _{_{2}}}n \Gamma
- \left(\frac{\mu _{_{1}}}{T _{_{1}}} - \frac{\mu _{_{2}}}{T _{_{2}}}
\right)
n _{_{1}}\Gamma _{_{1}}
+ \left(\frac{1}{T _{_{1}}} - \frac{1}{T _{_{2}}}\right)u _{a}
t _{_{1}}^{a} {\mbox{ .}}
\label{28}
\end{equation}
The condition $S ^{a}_{;a} = 0$ requires the well-known
equilibrium
conditions (see, e.g., \cite{Bern} (chapter 5))
\begin{equation}
\mu _{_{1}} = \mu _{_{2}}   {\mbox{ , }}
T _{_{1}} = T _{_{2}} {\mbox{ , }}
\label{29}
\end{equation}
as well as $\Gamma = 0$.

From now on we assume that the source terms on the right-hand side  
of eq.(\ref{13}) cancel among themselves, i.e., that the entropy
per particle of each of the components is preserved.
The particles decay or come into being with a fixed
entropy $s _{_{A}}$.
This `isentropy' condition
amounts to the assumption that the particles at any stage are
amenable to a perfect fluid description.
With $\dot{s}_{_{A}} = 0$, via eq.(\ref{13}) equivalent to
\begin{equation}
u _{a} t ^{a}_{_{A}} = \left(\rho _{_{A}} + p _{_{A}}\right)
\Gamma _{_{A}}
{\mbox{ , }}
\label{30}
\end{equation}
the expression (\ref{17}) for the temperature behaviour and
the expression (\ref{28}) for the entropy production
density simplify considerably.
The condition (\ref{30}) establishes a relation between
the source terms in the balances (\ref{4}) and (\ref{10}) which  
originally
are independent quantities.

Combining eqs.(\ref{21}) and (\ref{30}), one has
\begin{equation}
u _{a} t ^{a}_{_{1}} = \left(\rho _{_{1}} + p _{_{1}}\right)
\Gamma _{_{1}}
= - u _{a} t ^{a}_{_{2}} = - \left(\rho _{_{2}} + p _{_{2}}\right)
\Gamma _{_{2}}
{\mbox{ , }}
\label{31}
\end{equation}
which provides us with a relation between the rates
$\Gamma _{_{1}}$ and
$\Gamma _{_{2}}$:
\begin{equation}
\Gamma _{_{2}} = - \frac{\rho _{_{1}} + p _{_{1}}}
{\rho _{_{2}} + p _{_{2}}}
\Gamma _{_{1}}
{\mbox{ .}}
\label{32}
\end{equation}
Inserting the last relation into equation (\ref{24})
yields
\begin{equation}
n \Gamma = n _{_{1}}\Gamma _{_{1}} h _{_{1}}
\left[\frac{1}{h _{_{1}}}
- \frac{1}{h _{_{2}}}\right] {\mbox{ .}}
\label{33}
\end{equation}
The quantities $h _{_{A}} \equiv \left(\rho _{_{A}} + p _{_{A}}
\right)/ n _{_{A}}$ are the enthalpies per particle.
Total particle number conservation, i.e., $\Gamma = 0$
is only possible if the enthalpies per particle of both
components coincide, i.e., for $h _{_{1}} = h _{_{2}}$.

With the relations (\ref{30}) and (\ref{33}) the entropy production 
density (\ref{28}) becomes
\begin{equation}
S ^{a}_{;a} = \left(\rho _{_{1}} + p _{_{1}}\right)
\left[\frac{n _{_{1}} s _{_{1}}}{\rho _{_{1}} + p _{_{1}}}
- \frac{n _{_{2}} s _{_{2}}}{\rho _{_{2}} + p _{_{2}}}\right]
\Gamma _{_{1}}
= n _{_{1}}\Gamma _{_{1}} h _{_{1}}
\left[\frac{s _{_{1}}}{h _{_{1}} }
- \frac{ s _{_{2}}}{h _{_{2}}}\right]
{\mbox{ .}}
\label{34}
\end{equation}
We emphasize that acording to the equations of state
(\ref{14}) and (\ref{15}) the quantities $\rho _{1}$,
$p _{_{1}}$ and $s _{_{1}}$ depend on $T _{_{1}}$, while
$\rho _{_{2}}$, $p _{_{2}}$ and $s _{_{2}}$ depend on $T _{_{2}}$.
In general, we have $T _{_{1}} \neq T _{_{2}}$.

From the expression (\ref{34}) it is obvious that a decay of fluid $1$ 
into fluid $2$ particles is accompanied by a nonvanishing
entropy production unless
$s _{_{1}}/s _{_{2}} = h _{_{1}}/h _{_{2}}$.
For $\mu _{_{1}} = \mu _{_{2}} = 0$ the latter condition reduces
to $T _{_{1}} = T _{_{2}}$.
This situation corresponds to the case that all the entropy
of fluid $1$ is transferred into that of fluid $2$.
On this basis one usually discusses, e.g., the cosmological
electron-positron annihilation at an energy scale
of about 1 MeV (\cite{KoTu,Bern,Boe,Pots}.)

\section{The effective one-temperature model}
There exists a more familiar, alternative description of
a two-component fluid close to equilibrium, that is
based on a single Gibbs-equation for the system as a whole:
\begin{equation}
T \mbox{d}s = \mbox{d} \frac{\rho}{n} + p \mbox{d} \frac{1}{n}
- \left(\mu _{_{1}} - \mu _{_{2}}\right) \mbox{d} \frac{n _{_{1}}}{n} 
{\mbox{ , }}
\label{35}
\end{equation}
where $s$ is the entropy per particle.
The temperature $T$ is the equilibrium temperature of the
whole system.
The temperatures $T _{_{1}}$ and $T _{_{2}}$ of the previous
section do not appear as variables in the present effective
one-temperature description.
An (approximate) equilibrium for the entire system is assumed
to be established through the interactions between the
subsystems on the right-hand side of the balances (\ref{10}) and  
(\ref{11}).
Furthermore,
we assume that analogous to the equations of state
(\ref{14}) and (\ref{15})
the cosmic fluid as a whole is characterized by equations
of state
\begin{equation}
p = p\left(n, n _{_{1}}, T\right)
\label{36}
\end{equation}
and
\begin{equation}
\rho = \rho \left(n, n _{_{1}}, T\right){\mbox{ , }}
\label{37}
\end{equation}
where $p$ is the equilibrium pressure and $\rho$ is the energy
density of the system as a whole.
From the special case of a mixture of radiation and
nonrelativistic matter
discussed below eq.(\ref{17}) it is obvious that the overall
quantities $p$ and $\rho $ depend on two independent number
densities which we have chosen to be
$n$ and $n _{_{1}}$.
As long as the pressures are those for classical gases, i.e.,
$p _{_{A}} = n _{_{A}} T$, the equilibrium pressure $p$ of the system 
as a whole depends on $n = n _{_{1}} + n _{_{2}}$ only and the
separate dependence on $n _{_{1}}$  on the right-hand side of
eq.(\ref{36}) may be omitted.

If in the expressions
(\ref{25}) for the entropies per particle
the temperatures $T _{_{1}}$ and $T _{_{2}}$ are identified among
themselves and with $T$, and $\rho \left(T\right) =
\rho _{_{1}}\left(T\right) +
\rho _{_{2}}\left(T\right)$ as well as  $p \left(T\right)
= p _{_{1}}\left(T\right) + p _{_{2}}\left(T\right)$ are used,
the description based on relation (\ref{35}) is consistent with
the description relying on the relations (\ref{12})
for $ns \left(T\right) =
n _{_{1}} s _{_{1}}\left(T\right)
+ n _{_{2}} s _{_{2}}\left(T\right)$.

The equilibrium temperature $T$ is {\it defined} by 
\cite{UI,ZMN}
\begin{equation}
\rho_{_{1}}\left(n_{_{1}},T_{_{1}}\right)
+ \rho_{_{2}}\left(n_{_{2}},T_{_{2}}\right) 
= \rho \left(n, n _{_{1}}, T\right)
{\mbox{ .}}
\label{38}
\end{equation}

As it was shown in \cite{ZMN} for $\Gamma = \Gamma _{_{1}}
= \Gamma _{_{2}} = 0$,
this generally  implies
\begin{equation}
p_{_{1}}\left(n_{_{1}},T_{_{1}}\right)
+ p_{_{2}}\left(n_{_{2}},T_{_{2}}\right) 
\neq p\left(n, n_{_{1}}, T\right)
{\mbox{ .}} \label{39}
\end{equation}
For perfect fluids $1$ and $2$ with separately conserved
particle numbers
the difference between both sides of the latter inequality is a 
viscous pressure.
Its existence is a consequence of
the different temperature evolution laws of the subsystems. 
As it is obvious from eq.(\ref{17})
the cooling rate $\dot{T}_{_{1}}/T _{_{1}}$ is different from
$\dot{T}_{_{2}}/T _{_{2}}$ even for $\Gamma _{_{1}} = \Gamma  
_{_{2}} = 0$
if the subsystems are governed by different equations of state.
The expansion of the Universe tends to increase the difference
between $T _{_{1}}$ and $T _{_{2}}$, i.e., to drive the system as
a whole away from equilibrium.
This manifests itself as a bulk viscous pressure \cite{ZMN}.

We shall show here that nonvanishing source terms $\Gamma _{_{A}}$
in the particle number balances (\ref{4}) give rise to a new
type of effective bulk pressure, i.e., to enlarged entropy production.  

In order to find an explicit description of this phenomenon we shall use 
the following line of arguing.
We {\it anticipate} that deviations from detailed balance, i.e.,
$\Gamma _{_{A}} \neq 0$, leading to $\Gamma \neq 0$ in general,
will generate an effective `reactive' bulk pressure $\pi _{_{react}}$. 
For the corresponding energy-momentum tensor of the system
as a whole we write from the outset
\begin{equation}
T ^{ik} = \rho u ^{i}u ^{k} + \left(p + \pi _{_{react}}\right) h ^{ik} 
{\mbox{ .}}
\label{40}
\end{equation}
To separate the reactive bulk pressure from
any other dissipative phenomenon, we
have ignored here the possibility of conventional bulk pressures,  
dealt with elsewhere \cite{ZMN} and
of nonvanishing heat fluxes and
shear stresses in inhomogeneous and anisotropic cosmological models. 
The relations $T ^{ik}_{\ ;k} = 0$ then
imply the energy balance
\begin{equation}
\dot{\rho} + \Theta \left(\rho + p + \pi _{_{react}}\right) = 0
\label{41}
\end{equation}
for the effective one-temperature description.
{\it Afterwards} we shall determine $\pi _{_{react}} $ by the  
consistency
requirement of the one-temperature description on the basis of
eqs.(\ref{35}), (\ref{4}),
(\ref{23}), and (\ref{41}), with the two-temperature description
based on eqs.(\ref{12}), (\ref{4}), (\ref{10}), (\ref{21}),
and (\ref{30}).
More specifically, $\pi _{_{react}}$ will be obtained by comparing
eq.(\ref{27}), leading to
the two-temperature expression
(\ref{34}) for the entropy production density with the
corresponding expression for $S ^{a}_{;a}$ to be calculated
below within the one-temperature picture.

From the Gibbs-equation (\ref{35}) one finds for the change
in the entropy per particle
\begin{equation}
n \dot{s} = - \frac{\Theta }{T}\pi _{_{react}} - \frac{\rho +  
p}{T}\Gamma
- \frac{n _{_{1}} n _{_{2}}}{n}\left(\frac{\mu _{_{1}}
- \mu _{_{2}}}{T}\right)
\left[\Gamma _{_{1}} - \Gamma _{_{2}}\right] {\mbox{ .}}
\label{42}
\end{equation}
Even for $\dot{s}_{_{1}} = \dot{s}_{_{2}} = 0$ we have
$\dot{s} \neq 0$ in general.
From $S ^{a} = nsu ^{a}$ the expression for the entropy
production density is
\begin{equation}
S ^{a}_{;a} = ns \Gamma + n \dot{s} {\mbox{ .}}
\label{43}
\end{equation}
Introducing here eq.(\ref{42}), we find
\begin{equation}
S ^{a}_{;a} = - \frac{\Theta}{T}\pi _{_{react}}  - \frac{n \mu}{T}\Gamma 
- \frac{n _{_{1}} n _{_{2}}}{n}\left(\frac{\mu _{_{1}}
- \mu _{_{2}}}{T}\right)
\left[\Gamma _{_{1}} - \Gamma _{_{2}}\right] {\mbox{ , }}
\label{44}
\end{equation}
where
\begin{equation}
\mu  = \frac{\rho + p}{n} - T s
\label{45}
\end{equation}
is the effective one-component chemical potential.

Applying in eq.(\ref{44}) the relation (\ref{32}) as well
as the decomposition
\begin{equation}
n \mu = n _{_{1}}\mu _{_{1}} + n _{_{2}}\mu _{_{2}}
\label{46}
\end{equation}
for the chemical potential, the entropy production density
may be written as
\begin{equation}
S ^{a}_{;a} = - \frac{\rho _{_{1}} + p _{_{1}}}{T}\Gamma _{_{1}}
\left(\frac{n _{_{1}} \mu _{_{1}}}{\rho _{_{1}} + p _{_{1}}}
- \frac{n _{_{2}} \mu _{_{2}}}{\rho _{_{2}} + p _{_{2}}}\right)
- \frac{\Theta }{T} \pi _{_{react}}
{\mbox{ .}}
\label{47}
\end{equation}
Here we used
$\rho \left(T\right) = \rho _{_{1}}\left(T\right)
+ \rho _{_{2}}\left(T\right)$ and
$p\left(T\right) = p _{_{1}}\left(T\right)
+ p _{_{2}}\left(T\right)$,
keeping in mind that all the quantities, including
those of the subsystems, depend on $T$, {\it not} on $T _{_{A}}$.   
There is only one temperature in this kind of description,
namely $T$.

The expression (\ref{47}) for the entropy production density
is a sum of terms each being a product of thermodynamical
`fluxes' and thermodynamical `forces'.
According to the standard phenomenological theory,
the requirement
$S ^{a}_{;a} \geq 0$ may be fulfilled by linear relations
between `fluxes' and `forces'. The new aspect here,
compared with the standard one is, that one has
two scalar fluxes, namely $\pi _{_{react}}$ and $\Gamma _{_{1}}$.
This will generally result in cross effects between
both phenomena.
A related discussion of chemical reactions in the
expanding Universe
was given in  \cite{Buch}.

Within the causal thermodynamics of irreversible
processes, both fluxes become dynamical degrees of
freedom on their own
and one  arrives at coupled evolution equations for
$\pi _{_{react}}$ and $\Gamma _{_{1}}$.
Investigations along these lines have been carried
out for a one-component system by Gariel and
Le Denmat \cite{GaLeDe}.

In the present paper we do not follow this kind of
arguing of the phenomenological theory.
Instead, we shall find an explicit expression for
$\pi _{_{react}}$ in terms of $\Gamma _{_{A}}$ with the help of
semiquantitative arguments.
The quantity connecting $\pi _{_{react}}$ and $\Gamma _{_{1}}$
represents a new kinetic coefficient, corresponding
to the mentioned cross effect.

Strictly speaking, the computation of $\pi _{_{react}}$ has to
come from kinetic theory,  which, however, is beyond
the scope of the present paper.
Our derivation may be regarded as intermediate between
the phenomenological and kinetic levels of description.

From eq.(\ref{27})
the  alternative expression for the entropy production
density $S ^{a}_{;a}$
within the two-temperature model becomes
\begin{equation}
S ^{a}_{;a} = n _{_{1}} s _{_{1}}\Gamma _{_{1}}
+ n _{_{2}} s _{_{2}}
\Gamma _{_{2}}
{\mbox{ , }}
\label{48}
\end{equation}
with $s _{_{1}} = s _{_{2}}\left(T _{_{1}}\right)$ and
$s _{_{2}} = s _{_{2}}\left(T _{_{2}}\right)$ from (\ref{25}),
where we assumed each of the subfluids to be perfect
on their own, i.e.,
$\dot{s}_{_{1}} = \dot{s}_{_{2}} = 0$.
The two-temperature
formula (\ref{48}) has to be consistent with
the one-temperature formula (\ref{47}).
This consistency requirement will be used to
determine $\pi _{_{react}}$.
The obvious difficulty one encounters here is that
one has to deal with three generally different temperatures:
The temperatures $T _{_{1}}$ and $T _{_{2}}$ of the
components $1$ and $2$ in formula (\ref{48})
and the equilibrium temperature $T$ of the system as a
whole in (\ref{47}).
The dynamics of $T _{_{1}}$ and $T _{_{2}}$ are given by eqs.(\ref{17}) 
with eqs.(\ref{30}).
In order to compare the expressions (\ref{47}) and (\ref{48})
for the entropy production we have to find a corresponding law
for $T$.
Such a law may be obtained via similar steps that led
us to the expression (\ref{17}).
There exists, however, the following complication.
Because of the additional dependence of $\rho $ on $n _{_{1}}$
one has now three partial derivatives of $\rho $:
$\left(\partial \rho / \partial T\right)_{n, n _{_{1}}}$,
$\left(\partial \rho / \partial n\right)_{T, n _{_{1}}}$, and
$\left(\partial \rho / \partial n _{_{1}}\right)_{n, T}$.
The requirement that $s$ is a state function now leads to
(\cite{ZiSpon,ZCQG})
\begin{equation}
\frac{\partial \rho }{\partial n} = \frac{\rho + p}{n}
- \frac{T}{n}\frac{\partial p}{\partial T}
- \frac{n _{_{1}}}{n} \left[\left(\mu _{_{1}} - \mu _{_{2}}\right)
- T \frac{\partial }{\partial T}\left( \mu _{_{1}} - \mu _{_{2}}
\right)\right] {\mbox{ , }}
\label{49}
\end{equation}
generalizing relation (\ref{16}),
and the additional relation
\begin{equation}
\frac{\partial \rho }{\partial n _{_{1}}} =
\mu _{_{1}} - \mu _{_{2}}
- T \frac{\partial }{\partial T}\left( \mu _{_{1}} - \mu _{_{2}}\right) 
{\mbox{ .}}
\label{50}
\end{equation}
Using the Gibbs-Duhem relations
\begin{equation}
\mbox{d} p _{_{A}} = n _{_{A}} s _{_{A}}
\mbox{d} T _{_{A}} + n _{_{A}}
\mbox{d}\mu _{_{A}}
\label{51}
\end{equation}
for $T _{_{A}} = T$ together with eq.(\ref{25}), one finds
\begin{equation}
\mu _{_{A}} - T
\frac{\partial \mu _{_{A}}}{\partial T} = \frac{\rho _{_{A}}}
{n _{_{A}}}
{\mbox{ .}}
\label{52}
\end{equation}
Consequently, the relations (\ref{49}) and (\ref{50}) may be
written as
\begin{equation}
\frac{\partial \rho }{\partial n} = \frac{\rho + p}{n}
- \frac{T}{n}\frac{\partial p}{\partial T}
- \frac{n _{_{1}}}{n} \left[\frac{\rho _{_{1}}}{n _{_{1}}}
- \frac{\rho _{_{2}}}{n _{_{2}}} \right]
\label{53}
\end{equation}
and
\begin{equation}
\frac{\partial \rho }{\partial n _{_{1}}} =
\frac{\rho _{_{1}}}{n _{_{1}}} - \frac{\rho _{_{2}}}{n _{_{2}}}
{\mbox{ , }}
\label{54}
\end{equation}
with $\rho _{_{1}} = \rho _{_{1}}\left(T\right)$ and
$\rho _{_{2}} = \rho _{_{2}}\left(T\right)$, since we are
within the one-temperature description.

Differentiating eq.(\ref{37}), using eqs.(\ref{4}), (\ref{23}),
(\ref{41}), (\ref{53}), and (\ref{54}),
we obtain
\begin{eqnarray}
\frac{\partial \rho }{\partial T} \dot{T} &=&
- \left(\Theta - \Gamma \right) T \frac{\partial p}
{\partial T}
- \left[\Theta \pi _{_{react}} + \Gamma \left(\rho + p\right)\right]
\nonumber\\
&& - \frac{n _{_{1}} n _{_{2}}}{n} \left(\frac{\rho _{_{1}}}{n _{_{1}}} 
- \frac{\rho _{_{2}}}{n _{_{2}}}\right)\left[\Gamma _{_{1}}
- \Gamma _{_{2}}\right]
{\mbox{ .}}
\label{55}
\end{eqnarray}
Applying here eq.(\ref{32}), following from
$\dot{s}_{_{1}} = \dot{s}_{_{2}} = 0$,
and restricting ourselves from now on to
classical fluids with
$p _{_{A}} = n _{_{A}} T$, we find the following
evolution law for the equilibrium temperature:
\begin{equation}
\frac{\dot{T}}{T} = - \left(\Theta - \Gamma\right)
\frac{\partial{p}}{\partial{\rho }}
- \frac{\Theta \pi _{_{react}}}{T \partial \rho / \partial T}
{\mbox{ , }}
\label{56}
\end{equation}
where the abbreviation
\begin{equation}
\frac{\partial{p}}{\partial{\rho}} \equiv
\frac{\partial p / \partial T}{\partial \rho / \partial T}
\label{57}
\end{equation}
was used. \\
The temperature law (\ref{56}) is a new result.
It describes the behaviour of the equilibrium temperature of
an interacting and reacting two-fluid mixture with
nonconserved total particle number.
It is different in its structure from the laws (\ref{17})
with (\ref{30})
because of the $\pi _{_{react}}$ term.
The result (\ref{56}) differs also from previous one-component
approaches to particle production in the expanding Universe
\cite{LiGer,ZPRD,ZP1}.
The point here is that the `isentropy' conditions
$\dot{s}_{_{A}} = 0$, equivalent to relation (\ref{30}) of the
present paper are physically different from the
case of `adiabatic' particle production in
\cite{LiGer,ZPRD,ZP1}, characterized by $\dot{s} = 0$.
We conclude that `adiabatic' particle production in
the sense of \cite{LiGer,ZPRD,ZP1}  is not consistent
with the requirement of a perfect fluid behaviour of the
subsystems in a reacting two-component mixture.
\section{The effective viscous pressure}
With the formulae (\ref{17}) and (\ref{56}) we know the evolution
laws for all three temperatures.
We are now prepared to compare the expressions
(\ref{47}) and (\ref{48}) for the entropy
production densities.
To that purpose we write $s _{_{1}}
= s _{_{1}}\left(T _{_{1}}\right)$ and
$s _{_{2}} = s _{_{2}}\left(T _{_{2}}\right)$  as ($A = 1, 2$)
\begin{equation}
s _{_{A}}\left(T _{_{A}}\right)
= s _{_{A}}\left(T\right) + \Delta s _{_{A}}
{\mbox{ .}}
\label{58}
\end{equation}
Using this splitting in eq.(\ref{48}) and comparing with
eq.(\ref{47}) yields
\begin{equation}
- \frac{\Theta }{T}\pi _{_{react}} = \Gamma _{_{1}} n _{_{1}}
\Delta s _{_{1}}
+ \Gamma _{_{2}} n _{_{2}} \Delta s _{_{2}}  {\mbox{ .}}
\label{59}
\end{equation}
Since according to eqs.(\ref{17}) and (\ref{56})
the temperatures $T _{_{A}}$
and $T$ are different,  the quantities
$\Delta s _{_{A}} = s _{_{A}}\left(T _{_{A}}\right) -
s _{_{A}}\left(T\right)$ will be nonvanishing in general.
However, as long as the detailed balance relations
$\Gamma _{_{A}} = 0$ are fulfilled, $\pi _{_{react}}$ vanishes
even for $\Delta s _{_{A}} \neq 0$.
(We recall that this case leads to conventional
bulk stresses, discussed elsewhere \cite{ZMN}).

In order to find an explicit expression for $\pi _{_{react}}$
we use the following
heuristic mean free time argument (cf. \cite{ZMN}).
Let $\tau$ be the characteristic mean free time
for the interaction
between both components, generally symbolized
by $t _{_{A}}^{a}$ in eq.(\ref{9}).  
The time $\tau$ is assumed to be much larger
than any characteristic interaction time within
each of the fluids. Consequently, the latter may
be regarded as perfect on time scales of the order of $\tau$.
The interaction between the fluids is pictured by
`collisional'  events, where $\tau$ plays the role
of a mean free time between these `collisions'.
During the time interval $\tau$, i.e., between
subsequent interfluid interaction events, both
components then evolve according to their internal
perfect fluid dynamics, given by eqs.(\ref{4}), (\ref{10})
and (\ref{17}) with condition (\ref{30}).

Let us now assume that at $t = t _{_{0}}$ there exist
deviations from
detailed balance due to the circumstance that one of
the components, say fluid $1$, becomes nonrelativistic
and starts to decay into the relativistic fluid $2$.
Let us further assume that both components were in
thermal equilibrium at
$t = t _{_{0}}$ due to their mutual interaction, i.e.,
$T _{_{1}}\left(t _{_{0}}\right) =
T _{_{2}}\left(t _{_{0}}\right) = T \left(t _{_{0}}\right)$,
equivalent to
$\Delta s _{_{1}}\left(t _{_{0}}\right)
= \Delta s _{_{2}}\left(t _{_{0}}\right) =
\pi _{_{react}}\left(t _{_{0}}\right) = 0$.
Until a subsequent interaction event after a
mean free time $\tau $
between the `collisions' the components evolve
effectively free due to their perfect fluid dynamics.
According to the temperature laws (\ref{17}) and
(\ref{56}), the temperatures $T _{_{1}}$, $T _{_{2}}$
and $T$  evolve differently for $t > t _{_{0}}$.
At the time $t$ with $t - t _{_{0}} \leq \tau $ we
will have $T \neq T _{_{A}}$ in general with $T _{_{1}} \neq T _{_{2}}$. 
Consequently, the quantities $\Delta s _{A}$ will no
longer vanish at the time $t$.
Nonvanishing values of $\Delta s _{_{A}}$, however, together with
$\Gamma _{_{A}} \neq 0$,
will give rise to an effective bulk pressure according to
eq.(\ref{59}).

Up to first order in the temperature deviations we may write
\begin{equation}
\Delta s _{_{A}} = \left(T _{_{A}} - T\right)
\left(\frac{\partial s _{_{A}}}{\partial T}\right)_{n _{_{A}}} + ... 
{\mbox{ .}}
\label{60}
\end{equation}
Applying the general formula
\begin{equation}
\frac{\partial s _{_{A}}}{\partial T} = \frac{1}{n _{_{A}}
T _{_{A}}}
\frac{\partial \rho _{_{A}}}{\partial T _{_{A}}} {\mbox{ , }}
\label{61}
\end{equation}
that follows from the Gibbs-Duhem relation (\ref{51}), we get
\begin{equation}
- \Theta \pi _{_{react}} = n _{_{1}} \Gamma _{_{1}}
c _{v}^{^{\left(1 \right)}}
\left(T _{_{1}} - T\right)
+ n _{_{2}} \Gamma _{_{2}} c _{v}^{^{\left(2 \right)}}
\left(T _{_{2}}
- T\right)
{\mbox{ , }}
\label{62}
\end{equation}
where
\begin{equation}
c _{v}^{^{\left(A \right)}} \equiv \frac{1}{n _{_{A}}}
\frac{\partial \rho _{_{A}}}{\partial T}
\label{63}
\end{equation}
are the specific heats of the components.
Introducing the relation (\ref{32}) between the rates
$\Gamma _{_{1}}$ and $\Gamma _{_{2}}$, eq.(\ref{62})
may be written as
\begin{eqnarray}
- \Theta \pi _{_{react}}&=& \Gamma _{_{1}} \left(\rho _{_{1}}
+ p _{_{1}}\right)
\left[\frac{n _{_{1}} c _{v}^{^{\left(1 \right)}}
\left(T _{_{1}} - T\right)}
{\rho _{_{1}} + p _{_{1}}}
- \frac{n _{_{2}} c _{v}^{^{\left(2 \right)}}
\left(T _{_{2}} - T\right)}
{\rho _{_{2}} + p _{_{2}}}\right] \nonumber\\
&=& n _{_{1}}\Gamma _{_{1}} h _{_{1}}
\left[\frac{ c _{v}^{^{\left(1 \right)}}
\left(T _{_{1}} - T\right)}
{h _{_{1}}}
- \frac{c _{v}^{^{\left(2 \right)}}
\left(T _{_{2}} - T\right)}
{h _{_{2}}}\right]
{\mbox{ .}}
\label{64}
\end{eqnarray}
We recall that this reactive bulk pressure is a cross effect
between the deviation from detailed balance, represented by
$\Gamma _{_{1}} \neq 0$, and the deviation from thermal
equilibrium, i.e.,
$T _{_{1}} \neq T _{_{2}} \neq T$.
The conventional bulk pressure is only due to the
different cooling rates.
Here, both deviations from the detailed balance and
from the thermal equilibrium are necessary.

Let again component $1$ be the decaying component
(i.e., $\Gamma _{_{1}} < 0$)
with an equation of state close to that for matter.
Because of
$\partial p _{_{1}}/ \partial \rho _{_{1}}
\geq \partial p/ \partial \rho
\geq \partial p _{_{2}}/ \partial \rho _{_{2}} $,
the temperature
$T _{_{1}}$ cools off faster than the equilibrium
temperature $T$, while
$T _{_{2}}$ decreases slower than $T$ [cf.(\ref{17}),
(\ref{56})].
Consequently, we expect $T _{_{1}} < T$ and $T _{_{2}} > T$.
In total, the result according to eq.(\ref{64})
will be a negative bulk pressure corresponding to a
positive effective bulk viscosity, in accordance with
the second law of thermodynamics.

Restricting ourselves to small temperature differences  we have
\begin{equation}
T _{_{A}} = T _{_{A}}\left(t _{_{0}}\right)
+ \left(t - t _{_{0}}\right)\dot{T}_{_{A}} + ....{\mbox{ , }}
\label{65}
\end{equation}
and
\begin{equation}
T  = T \left(t _{_{0}}\right)
+ \left(t - t _{_{0}}\right)\dot{T} + ....
{\mbox{ .}}
\label{66}
\end{equation}
Here, $\dot{T}_{_{A}}$ and $\dot{T}$ have to be evaluated at
$t = t _{_{0}}$.
While $\dot{T}_{_{A}}$ may be taken from eq.(\ref{17})
immediately, we have still to show that
$\dot{T}\left(t _{_{0}}\right)$ is consistent with
$\pi _{_{react}}\left(t _{_{0}}\right) = 0$.
This may be done by applying the relation (\ref{38})
that defines the equilibrium temperature $T$ at $t > t _{_{0}}$,
keeping in mind that the interaction between the fluids,
generally given by the source terms in the balances (\ref{10}) and
(\ref{11}), is modelled by `collisional' events,
with a mean free time $\tau $ between the `collisions'.
During the interval $\left(t - t _{_{0}}\right)
\leq \tau $, i.e., between subsequent interaction events,
both
components are effectively free, i.e.,
governed by their internal perfect
fluid
dynamics
according to eqs.(\ref{4}), (\ref{10}),
(\ref{17}), and (\ref{30}).  
Assume again that through the interaction an
element of the cosmic fluid
is in equilibrium at a proper time $t_{_{0}}$ at a
temperature 
$T\left(t_{_{0}}\right) = T_{_{1}}\left(t_{_{0}}\right) 
= T_{_{2}}\left(t_{_{0}}\right)$ with 
$p\left(t_{_{0}}\right) = p_{_{1}}\left(t_{_{0}}\right) + 
p_{_{2}}\left(t_{_{0}}\right)$. 
Here,  
$p\left(t_{_{0}}\right)$ and $p_{_{A}}\left(t_{_{0}}\right)$  
are shorts for 
$p\left[n\left(t_{_{0}}\right), T\left(t_{_{0}}\right)\right]$ 
and 
$p_{_{A}}\left[n_{_{A}}\left(t_{_{0}}\right),
T_{A}\left(t_{_{0}}\right)\right]$, respectively. 
Using the condition (\ref{38}) 
at the proper time 
$t$ up to first order in $t - t _{_{0}}$, i.e., with
\begin{equation}
\rho_{_{A}}\left(t\right) = 
\rho_{_{A}}\left(t_{_{0}}\right) + \left(t - t _{_{0}}\right)
\dot{\rho}_{_{A}}\left(t_{_{0}}\right) + ...\label{67}
\end{equation} 
and 
\begin{equation}
\rho\left(t\right) = 
\rho\left(t_{_{0}}\right) + \left(t - t _{_{0}}\right)
\dot{\rho}\left(t_{_{0}}\right) + ... \label{68}
\end{equation}
where 
$\rho_{_{A}}\left(t_{_{0}}\right) \equiv 
\rho_{_{A}}\left[n_{_{A}}\left(t_{_{0}}\right),
T_{_{A}}\left(t_{_{0}}\right)\right]$ 
and 
$\rho\left(t_{_{0}}\right) \equiv 
\rho\left[n\left(t_{_{0}}\right),
n_{_{1}}\left(t_{_{0}}\right),
T\left(t_{_{0}}\right)\right]$, 
applying eq.(\ref{10})  
on the left-hand side of eq.(\ref{38})  and using the relations
(\ref{53})
and (\ref{54}) on its righ-hand side,
one finds
\begin{equation}
\dot{T}\left(t_{_{0}}\right)  = - T
\left(\Theta - \Gamma \right)
\frac{\partial p}{\partial \rho }
{\mbox{ , }}
\label{69}
\end{equation}
where all the quantities on the right-hand side of the last equation  
have to be taken
at the point $t = t_{_{0}}$. 
Comparing with the temperature law (\ref{56}),
this proves the consistency of our initial configuration with
$\pi _{_{react}}\left(t _{_{0}}\right) = 0$ for
$T _{_{1}}\left(t _{_{0}}\right) = T _{_{2}}\left(t _{_{0}}\right)
= T \left(t _{_{0}}\right)$, i.e., $\Delta s _{_{A}} = 0$
[cf. eq.(\ref{59})] and
$\Gamma _{_{A}}\left(t _{_{0}}\right) \neq 0$.

Applying now the evolution laws (\ref{17}) and (\ref{69})
at $t = t _{_{0}}$
we get for the first-order temperature differences
\begin{equation}
T _{_{A}} - T = - \Theta T \left(t - t _{_{0}}\right)
\left[\left(1 - \frac{\Gamma _{_{A}}}{\Theta }\right)
\frac{\partial p _{_{A}}}{\partial \rho _{_{A}}}
- \left(1 - \frac{\Gamma }{\Theta }\right)
\frac{\partial p}{\partial \rho }\right] + ....
{\mbox{ .}}
\label{70}
\end{equation}

The first-order approximation is valid as long as the
conditions \newline
$|\Theta - \Gamma _{_{A}}| \left(t - t _{_{0}}\right) \leq 1$
are fulfilled.
Recalling that an effective one-fluid description
requires $\tau < H ^{-1}$,
where $H = \Theta /3$ is the Hubble parameter, one may discuss three 
different cases. \\
(i) $|\Gamma _{_{A}}| ^{-1} < \tau < H ^{-1}$.
There is a considerable change of the particle numbers on
time scales of the order of the mean free time between
the interfluid interaction events.
\\
(ii) $\tau < H ^{-1} < |\Gamma _{_{A}}|^{-1}$.
Here, the characteristic times $|\Gamma _{A}|^{-1}$
for the particle number changes are larger than the Hubble time.
 \\
(iii) $\tau < |\Gamma _{A}|^{-1} < H ^{-1}$.
This case is intermediate between (i) and (ii).
The particle numbers change substantially during one Hubble
time but are approximately constant on scales of the order of
$\tau $.

Inserting the temperature differences (\ref{70}) into eq.(\ref{64}), 
the reactive viscous pressure may be written as
\begin{eqnarray}
\pi _{_{react}} &=& \Gamma _{_{1}} \eta  T
\frac{\partial \rho _{_{1}}}{\partial T}
\left[\left(1 - \frac{\Gamma _{_{1}}}{\Theta }\right)
\frac{\partial p _{_{1}}}{\partial \rho _{_{1}}}
- \left(1 - \frac{\Gamma }{\Theta }\right)
\frac{\partial p}{\partial \rho }\right] \nonumber\\
&& +
\Gamma _{_{2}} \eta   T
\frac{\partial \rho _{_{2}}}{\partial T}
\left[\left(1 - \frac{\Gamma _{_{2}}}{\Theta }\right)
\frac{\partial p _{_{2}}}{\partial \rho _{_{2}}}
- \left(1 - \frac{\Gamma }{\Theta }\right)
\frac{\partial p}{\partial \rho }\right]
{\mbox{ , }}
\label{71}
\end{eqnarray}
where
$\eta   \equiv t - t _{_{0}}$.
It can now be seen explicitly that $\pi _{_{react}}$ is always negative. 
For `ordinary' matter $\partial p_{_{A}}/\partial \rho_{_{A}}$
lies in the
range $1/3 \leq \partial p_{_{A}}/\partial \rho_{_{A}} \leq 2/3$.
The lower limit corresponds to
radiation, the upper one to matter. 
$\partial p/\partial \rho$ will take a
value intermediate between 
$\partial p_{_{1}}/\partial \rho_{_{1}}$ and 
$\partial p_{_{2}}/\partial \rho_{_{2}}$. 
The typical case is that massive particles decay into
relativistic ones.
Identifying again fluid $1$ with the massive decaying
component, we have $\Gamma _{_{1}} < 0$.
Furthermore,   
$\partial p/\partial \rho \leq  
\partial p_{_{1}}/\partial \rho_{_{1}}$ and 
$\partial p/\partial \rho \geq 
\partial p_{_{2}}/\partial \rho_{_{2}}$. 
Because of $\partial p _{_{1}}/\partial \rho _{_{1}}
> \partial p/\partial \rho $ one has
$\left(1 - \Gamma _{_{1}}/\Theta \right)
\partial p _{_{1}}/\partial \rho _{_{1}}
> \left(1 - \Gamma / \Theta \right)
\partial p/ \partial \rho $, since a negative
$\Gamma _{_{1}}$ enlarges the left-hand side of this
inequality and a positive $\Gamma $ diminishes its right-hand side  
In the second term one has
$\partial p _{_{2}}/\partial
\rho _{_{2}} < \partial p/\partial \rho $.
The relations (\ref{32}) and (\ref{33}) may be
combined into
\begin{equation}
\Gamma  = \frac{n _{_{2}}}{n}
\left[1 - \frac{h _{_{2}}}{h _{_{1}}}\right]
\Gamma _{_{2}} {\mbox{ .}}
\label{72}
\end{equation}
One has $\Gamma < \Gamma _{_{2}}$ for any
$\Gamma _{_{2}} > 0$.
It follows that the factor
$\left(1 - \Gamma _{_{2}}/\Theta \right)$
reduces the  $\partial p _{_{2}}/\partial \rho _{_{2}}$
term more than
$\left(1 - \Gamma / \Theta \right)$ reduces
the $\partial p/ \partial \rho $ term. Consequently,
$\left(1 - \Gamma _{_{2}}/\Theta \right)
\partial p _{_{2}}/\partial \rho _{_{2}}
< \left(1 - \Gamma / \Theta \right)\partial p/ \partial \rho $.
Both contributions to $\pi _{_{react}}$ in the expression (\ref{71}) 
are negative, i.e., the reactive bulk viscosity coefficient is
positive, as is required  by the second law of thermodynamics.
Inserting the expressions (\ref{32}) and (\ref{63}) into equation
(\ref{71}),
the reactive viscous pressure may be written as
\begin{eqnarray}
\pi _{_{react}}&=& \Gamma _{_{1}} \eta
n _{_{1}}h _{_{1}} T
\left[\frac{c _{v}^{^{\left(1 \right)}}}{h _{_{1}}}
\left( \frac{\partial{p _{_{1}}}}{\partial{\rho _{_{1}}}} -  
\frac{\partial{p}}{\partial {\rho }}\right)
- \frac{ c _{v}^{^{\left(2 \right)}}}{h _{_{2}}}
\left(\frac{\partial{p _{_{2}}}}{\partial{\rho _{_{2}}}}
- \frac{\partial{p}}{\partial{\rho}}\right)\right] \nonumber\\
&&
- \Gamma _{_{1}} \eta
\frac{\Gamma _{_{1}}}{\Theta }n _{_{1}}^{2}h _{_{1}}^{2} T
\left[\frac{c _{v}^{^{\left(1 \right)}}}{n _{_{1}}h _{_{1}}^{2}}
\left(\frac{\partial p _{_{1}}}{\partial \rho _{_{1}}}
- \frac{n _{_{1}}}{n} h _{_{1}}
\left(\frac{1}{h _{_{1}}}
- \frac{1}{h _{_{2}}}\right)
\frac{\partial p}{\partial \rho }\right) \right. \nonumber\\
 && \left. \mbox{\ \ \ \ \ \ \ \ }
+ \frac{c _{v}^{^{\left(2 \right)}}}{n _{_{2}}h _{_{2}}^{2}}
\left(\frac{\partial p _{_{2}}}{\partial \rho _{_{2}}}
- \frac{n _{_{2}}}{n} h _{_{2}}
\left(\frac{1}{h _{_{2}}}
- \frac{1}{h_{_{1}}}\right)
\frac{\partial p}{\partial \rho }\right)
\right]
{\mbox{ .}}
\label{73}
\end{eqnarray}
This is the
general formula for the reactive bulk pressure in a
two-component cosmological fluid for small
deviations from thermal equilibrium.
It is the main result of the paper.
By eq.(\ref{73}) the quantity $\pi _{_{react}}$ is given in terms of 
$\Gamma _{_{1}}$ or, via eq.(\ref{33}), in terms of the
production rate $\Gamma $.
This demonstrates explicitly that particle production
in a two-component mixture is equivalent to an
effective bulk pressure.
The existence of a formula like eq.(\ref{73}) may be
regarded as a semiquantitative justification for
using effective viscous pressures in modelling
particle creation processes on a phenomenological level.
We recall that different from previous studies
(\cite{Calv,LiGer,ZP1,ZP2,ZP3,TZP})
the production process here is nonadiabatic.

In the following we shall apply formula (\ref{73})
to the
out-of-equilibrium decay of a nonrelativistic
fluid into a relativistic one.

\section{The out-of-equilibrium decay of heavy particles}
Decay processes of heavy particles are supposed to play
a role, e.g., during baryogenesis where hypothetical
heavy bosons decay into quarks and leptons.
Another application is the scalar field decay into
relativistic particles in the reheating phase of
standard inflationary scenarios (see, e.g., \cite{KoTu}).

We assume fluid $1$ to be described by the equations
of state for nonrelativistic matter, i.e.,
\begin{equation}
\rho _{_{1}} = n _{_{1}} m + \frac{3}{2} n _{_{1}} T _{_{1}}{\mbox{  
, }}
\mbox{\ \ \ \ \ }
p _{_{1}} = n _{_{1}} T {\mbox{ , }}
 \mbox{\ \ \ \ \ }
m \gg T {\mbox{ , }}
\label{104}
\end{equation}
while fluid $2$ is a classical relativistic gas:
\begin{equation}
\rho _{_{2}} = 3 n _{_{2}}T _{_{2}}{\mbox{ , }}
\mbox{\ \ \ \ \ \ \ }
p _{_{2}} = n _{_{2}} T _{_{2}} {\mbox{ .}}
\label{105}
\end{equation}
The corresponding enthalpies are
\begin{equation}
h _{_{1}} = m + \frac{5}{2}T _{_{1}}{\mbox{ , }}
\mbox{\ \ \ \ \ \ \ }
h_{_{2}} = 4 T _{_{2}} {\mbox{ , }}
\label{106}
\end{equation}
and the specific heats
\begin{equation}
c _{v}^{^{\left(1 \right)}} = \frac{3}{2} {\mbox{ , }}
\mbox{\ \ \ \ \ \ }
c _{v}^{^{\left(2 \right)}} = 3 {\mbox{ .}}
\label{107}
\end{equation}
The expressions for $\partial p _{_{A}}/ \partial \rho _{_{A}}$ and 
$\partial p / \partial \rho $ become
\begin{equation}
\frac{\partial p _{_{1}}}{\partial \rho _{_{1}}}
= \frac{2}{3}{\mbox{ , }}
 \mbox{\ \ \ \ \ \ }
\frac{\partial p _{_{2}}}{\partial \rho _{_{2}}}
= \frac{1}{3} {\mbox{ , }}
\mbox{\ \ \ \ \ \ }
\frac{\partial p }{\partial \rho } =
\frac{2}{3} \frac{n _{_{1}} + n _{_{2}}}{n _{_{1}}
+ 2 n _{_{2}}} {\mbox{ .}}
\label{108}
\end{equation}
Inserting eqs.(\ref{104})-(\ref{108}) into
eq.(\ref{73}) and taking into account
$m \gg T$ we obtain
\begin{equation}
\pi _{_{react}} = \frac{n _{1} m}{4} \frac{n _{_{1}}}{n _{_{1}}
+ 2 n _{_{2}}}
\left[1 - \frac{n _{_{1}} m}{4 n _{_{2}} T}
\frac{\Gamma _{_{1}}}{\Theta }\right]
\Gamma _{_{1}} \eta  {\mbox{ .}}
\label{109}
\end{equation}
Since $\rho + p = \rho _{_{1}} + \rho _{_{2}} + p _{_{1}} + p _{_{2}} 
\approx n _{_{1}} m + 4 n _{_{2}} T $, the ratio
$\pi _{_{react}}/ \left(\rho + p\right)$
is
\begin{equation}
\frac{\pi _{_{react}} }{\rho + p} = \frac{1}{4}
\frac{n _{_{1}} m}{n _{_{1}} m + 4 n _{_{2}} T}
\frac{n _{_{1}}}{n _{_{1}} + 2 n _{_{2}}}
\left[1 + \frac{n _{_{1}} m}{4 n _{_{2}} T}
\frac{|\Gamma _{_{1}}|}{\Theta }\right]
\Gamma _{_{1}} \eta   {\mbox{ .}}
\label{110}
\end{equation}
In order to appropriately apply this formula we have to take into
account the limitations following from the linear order restrictions 
(\ref{65})
and (\ref{66}) for the temperature evolution laws.
From eq.(\ref{17}) with (\ref{31}) we find the linear relations
(\ref{65}) and (\ref{66}) to make sense
for
\begin{equation}
|\Theta - \Gamma _{_{A}}| \left(t - t _{_{0}} \right) \leq 1 \ .
\label{111}
\end{equation}
To be cosmologically relevant, we expect $|\Gamma _{_{1}}|$ to be
of the order of $\Theta $, i.e.,
$-\Gamma _{_{1}} = |\Gamma _{_{1}}| = \alpha \Theta $, where $\alpha$ 
is a positive constant of the order $1$.
For component $1$ the linear evolution law is valid up to time  
intervals
\begin{equation}
\left(t - t _{_{0}} \right) \sim
\frac{1}{\Theta + |\Gamma _{_{1}}|}
= \frac{\alpha}{\alpha + 1}|\Gamma _{_{1}}|^{-1} \ .
\label{112}
\end{equation}
Assuming furthermore the nonrelativistic component $1$ to dominate
initially, it follows from eq.(\ref{32}) that $\Gamma _{_{2}} >  
|\Gamma _{_{1}}|$, such that
$\Gamma _{_{2}} > \Theta $ is possible.
In the latter case combination of (\ref{112}) with (\ref{111}) for  
$A = 2$ yields
\begin{equation}
\Gamma _{_{2}} \sim \frac{\alpha + 2}{\alpha} |\Gamma _{_{1}}| \ .
\label{113}
\end{equation}
Comparing the last expression with eq.(\ref{32}) we get, by using
the equations
of state (\ref{104}) and (\ref{105}),
\begin{equation}
\rho _{_{1}} \sim \frac{4}{3}\frac{\alpha + 2}{\alpha}
\rho _{_{2}}\ .
\label{114}
\end{equation}
The limitations of the linear approximation allow the energy  
density of the matter component to be larger than that of the  
radiation component by the factor $4 \left(\alpha + 2  
\right)/\left(3 \alpha \right)$.
For $\alpha \approx 1$ the ratio $\rho _{_{1}}/ \rho _{_{2}} \approx 4$ 
is allowed, while for $\alpha \approx 1/4$ we have
$\rho _{_{1}}/\rho _{_{2}} \approx 12$.
Since eq.(\ref{114}) is equivalent to
\begin{equation}
\frac{n _{_{1}}m}{4 n _{_{2}}T} \approx \frac{\alpha + 2}{\alpha} \ ,
\label{115}
\end{equation}
we find from eq.(\ref{109})
\begin{equation}
\pi _{_{react}}= - \frac{n _{_{1}}m}{4}
\frac{n _{_{1}}}{n _{_{1}} + 2 n _{_{2}}}
\left(\alpha + 3 \right)\alpha \eta \Theta \ ,
\label{116}
\end{equation}
corresponding to an effective bulk viscosity coefficient
$\zeta _{_{react}}$
\begin{equation}
\zeta _{_{react}} = \eta \frac{n _{_{1}}m}{4}
\frac{n _{_{1}}}{n _{_{1}} + 2 n _{_{2}}}
\alpha \left(\alpha + 3 \right)\ .
\label{117}
\end{equation}
It is interesting to compare this bulk viscosity coefficient with  
the conventional one, characterizing the viscous pressure in a  
two-fluid
system with conserved particle numbers.
In general, the latter is given by \cite{ZMN}
\begin{equation}
\zeta = - \tau T \frac{\partial{\rho }}{\partial{T}}
\left(\frac{\partial{p _{_{2}}}}{\partial{\rho _{_{2}}}}
- \frac{\partial{p}}{\partial{\rho }} \right)
\left(\frac{\partial{p _{_{1}}}}{\partial{\rho _{_{1}}}}
- \frac{\partial{p}}{\partial{\rho }} \right)\ ,
\label{118}
\end{equation}
which, for matter dominance, coincides in good approximation with
Weinberg's expression for the bulk viscosity in radiative hydrodynamics 
\cite{Wein,ZMN}.
With the equations of state (\ref{104}) and (\ref{105}) the expression 
(\ref{118}) reduces to
\begin{equation}
\zeta =   \frac{\tau }{3}n _{_{2}} T
\frac{n _{_{1}}}{n _{_{1}} + 2 n _{_{2}}}\ .
\label{119}
\end{equation}
Assuming now $\eta \approx \tau $ (see the discussion below  
eq.(\ref{70}))
we find
\begin{equation}
\frac{\zeta _{_{react}}}{\zeta } \approx 3
\frac{n _{_{1}}m}{4 n _{_{2}}T}\alpha \left(\alpha + 3 \right) \ ,
\label{120}
\end{equation}
and, with eq.(\ref{115}),
\begin{equation}
\frac{\zeta _{_{react}}}{\zeta } \approx 3
\left(\alpha + 2 \right) \left(\alpha + 3 \right) \ .
\label{121}
\end{equation}
For $\alpha \approx 1$ one has $\zeta _{_{react}} \approx 36 \zeta $, 
while for $\alpha \approx 1/4$ the quantity
$\zeta _{_{react}}$ is  more than $20$ times larger than $\zeta $.
{\it The reactive bulk viscosity exceeds the conventional
one by more than one order of magnitude and, consequently,
provides the dominating contribution to the bulk viscous part of
the entropy production
during the decay process}.

\section{Conclusions and outlook}
Particle number nonconserving reactions between
two cosmological fluids, each of them perfect on its own,
necessarily imply
the existence of entropy producing reactive bulk stresses.
This result of the present paper provides a semiquantitative
justification for the frequently used approach of
regarding particle creation processes as
phenomenologically equivalent to effective viscous pressures.
Different from most of the previous applications,
the particle production turned out to be `nonadiabatic'
in the present case, i.e., the entropy per particle of the cosmic
fluid as a whole does not remain constant.
For small deviations from thermal equilibrium we derived
a general formula for the corresponding new kinetic coefficient.
The latter was explicitly evaluated for the
out-of-equilibrium decay of nonrelativistic
fluid particles into relativistic ones.
The reactive bulk viscosity coefficient  in this case was shown
to be larger than the conventional bulk viscosity by more than
one order of magnitude. \\
Our results indicate that bulk pressures are a general
cosmological phenomenon.
While not taken into account in the standard perfect fluid approaches, 
one may expect them to have considerably influenced the
evolution of the early universe.
Of special interest in this respect are far-from-equilibrium situations. 
The existence of a new type of bulk pressure in addition to the
`conventional' one may shed new light, e.g., on the question whether 
bulk pressures via their backreaction on the cosmological dynamics
might give rise to a phase of inflationary growth of the cosmic
scale factor (see \cite{RMCQG,ZPRD,ZTP} and references therein). \\
An out-of-equilibrium period that recently has attracted much attention 
(see, e.g., \cite{Boya,Bberger}) is the reheating phase of  
inflationary universe models.
On the basis of the temperature evolution law (\ref{56})
it was shown to be possible to
interrelate the description of nonequilibrium
processes within the causal, second order M\"uller-Israel-Stewart  
theory with particle creation during the `preheating'  stage  
\cite{ZPM}.
A further subject of interest in this connection may be the `reheating' 
from primordial black hole evaporation \cite{Ga-Be}. \\

\ \\
{\bf Acknowledgement}\\
\ \\
This paper was supported by the Deutsche Forschungsgemeinschaft.
I am grateful to Diego Pav\'{o}n and Thomas Buchert for discussions  
and remarks. \\
\ \\
\begin{small}

\end{small}

\end{document}